\documentclass[12pt]{article}
\usepackage{amsmath}
\usepackage{latexsym}
\usepackage{a4wide}
\usepackage{bbm}
\usepackage{epsfig}
\newcommand{\I}{\text{i}}
\newcommand{\E}{\text{e}}
\newcommand{\tr}{\text{tr}}
\newcommand{\Tr}{\text{Tr}}
\newcommand{\re}[1]{(\ref{#1})}
\newcommand{\Nc}{N_{\text{c}}}
\newcommand{\meff}{m_{\text{eff}}}
\newcommand{\Geff}{\Gamma_{\text{eff}}}
\newcommand{\Leff}{{\cal L}_{\text{eff}}}
\newcommand{\Li}{\text{L{\small i}}}
\begin{document}
\title{\bf Effective action for the order parameter of the deconfinement
  transition of Yang-Mills theories}
\author{Holger Gies\\
  Institut f\"ur theoretische Physik, Universit\"at T\"ubingen,\\
  72076 T\"ubingen, Germany}
\maketitle
\begin{abstract}
The effective action for the Polyakov loop serving as an order
parameter for deconfinement is obtained in one-loop approximation to
second order in a derivative expansion. The calculation is performed
in $d\geq 4$ dimensions, mostly referring to the gauge group SU(2). The
resulting effective action is only capable of describing a
deconfinement phase transition for $d>d_{\text{cr}}\simeq
7.42$. Since, particularly in $d=4$, the system is strongly governed by
infrared effects, it is demonstrated that an additional infrared scale
such as an effective gluon mass can change the physical properties of
the system drastically, leading to a model with a deconfinement phase
transition. 
\end{abstract}

\section{Introduction}
As a prelude to a truly nonperturbative evaluation of the effective
action of Yang-Mills theory, the one-loop effective action with
all-order couplings to a specific background may provide a first
glance at the up to now unknown ground state of the theory. Since the
problem of confinement is supposed to be intimately related to the
quest for the ground state, it is elucidating to investigate the
response of several ``confining vacuum candidates'' on quantum
fluctuations -- even in a perturbative approximation. In this spirit,
e.g., the famous Savvidy model \cite{savv77}, which favors a covariant
constant magnetic field as ground state, has given rise to much
speculation on the nature of the vacuum.

Since a useful description of confinement and the ground state should
also exhibit the limits of their formation, it is natural to perform a
study at finite temperature where a transition to a deconfined phase
is expected (as is observed on the lattice). An order parameter for
the deconfinement phase transition in pure gauge theory is given by
the Polyakov loop \cite{poly78,suss79}, i.e., a Wilson line closing
around the compactified Euclidean time direction:
\begin{equation}
L(x) =\frac{1}{\Nc} \tr\, \text{T}\, \exp \left( \I g
  \int\limits_0^\beta dx_0\, \mathsf{A}_0(x_0,x) \right), \label{1}
\end{equation}
where the period $\beta =1/T$ is identified with the inverse
temperature of the ensemble in which the expectation value of $L$ is
evaluated. T denotes time ordering, $\Nc$ the number of colors, and
$\mathsf{A}_0$ is the time component of the gauge field. The negative
logarithm of the Polyakov loop expectation value can be interpreted as
the free energy of a single static color source living in the
fundamental representation of the gauge group \cite{svet86}. In this
sense, an infinite free energy associated with confinement is
indicated as $\langle L\rangle \to 0$, whereas $\langle L\rangle\neq
0$ signals deconfinement.

Moreover, $\langle L\rangle$ measures whether center symmetry, a
discrete symmetry of Yang-Mills theory, is realized by the ensemble
under consideration \cite{svet86}. Gauge transformations which differ
at $x_0=0$ and $x_0=\beta$ by a center element of the gauge group
change $L$ by a phase $\E^{2\pi\I n/\Nc}$, $n$ integer (but leave the
action invariant); this implies that a center-symmetric ground state
automatically ensures $\langle L\rangle=0$, whereas deconfinement
$\langle L\rangle\neq 0$ is related to the breaking of this symmetry.

Therefore, the effective action governing the behavior of $L$ is of
utmost importance, because it determines the state of the theory at a
given set of parameters, such as temperature, fields, etc. While this
scenario has successfully been established in lattice formulations
\cite{latt}, several perturbative continuum investigations have led to
various results. In the continuum, it is convenient to work with the
``Polyakov gauge'', which rotates the zeroth component $\mathsf{A}_0$
of the gauge field into the Cartan subalgebra of SU($\Nc$),
$\mathsf{A}_0\to A_0$ (cf. Eq. \re{2} below); 
furthermore, the condition $\partial_0 A_0=0$ is imposed. Then, if the
$A_0$ ground state of the system is known, $L$ can immediately be read
off from Eq.~\re{1}, which suggests calculating the effective action
of an time-independent $A_0$ background field. 

Several one-loop calculations exist in the literature: considering a
pure constant $A_0$ background, Weiss \cite{weis81} obtained an
effective potential for the Polyakov loop preferring only the
center-asymmetric ground states, i.e., the deconfined phase (see
Eq.~\re{weiss} below). Combining the Savvidy model of a covariant
constant magnetic background field with the Polyakov loop background
$A_0$, Starinets, Vshivtsev and Zhukovskii \cite{star94} as well as
Meisinger and Ogilvie \cite{meis97} were able to demonstrate the
existence of a confining center symmetric minimum for $\langle
L\rangle$ at low temperature with a transition to the broken,
i.e., deconfined phase for increasing temperature. However, this model
still suffers from the instabilities caused by the gluon spin coupling
to the magnetic field \cite{meis97}, a problem also plaguing the
Savvidy model \cite{niel78}, although an additional $A_0$ background
can in principle remove the problematic tachyonic mode in the gluon
propagator for certain values of $gA_0$ and $T$. Perhaps the most
promising approach was explored by Engelhardt and Reinhardt
\cite{enge98}, who considered a spatially varying $A_0$ field and
evaluated the effective action for $A_0$ in a gradient expansion to
second order in the derivatives. The resulting action exhibits both
phases, confinement and deconfinement, depending on the value of the
temperature; in particular at low temperature, the spatial
fluctuations of the Polyakov loop lower the action when fluctuating
around the center symmetric (confining) phase. The main drawback of
this model is its nonrenormalizability. An explicit cutoff dependence
remains; gauge and Lorentz invariance have been broken explicitly
during the calculation. Nevertheless, the main lesson to be learned is
that spatial variations of the Polyakov loop have to be taken into
account while searching for an effective potential of the order
parameter for the deconfinement phase transition.

The present work is devoted to an investigation of the Polyakov loop
potential partly in the spirit of \cite{enge98}; however, the treatment
of the quantum fluctuations, the calculational techniques, and finally
the results are quite different. In particular, we employ the
background field method to keep track of the symmetries of the
functional integral \cite{abbo81}. Unfortunately, the results are not
as promising as those found in \cite{enge98}, since the simple
picture for the deconfinement phase transition is not visible in the
most stringent version of the model. 

The paper is organized as follows: in Sec. \ref{model}, we define the
model, clarify our notations, and perform a first analysis of possible
scenarios. Section \ref{calculus} outlines the calculation of the
effective action to one loop using the proper-time method,
particularly emphasizing the subtleties of the present problem; we
work in $d\geq 4$ dimensions with gauge group SU($\Nc$). The
implications of our results are discussed in Sec.~\ref{analysis} for
SU(2); therein it is pointed out that the main features of the model
depend strongly on the treatment of the infrared sector. Section
\ref{gluonmass} briefly demonstrates the latter point by introducing an
additional infrared scale ``by hand'' (gluon mass), which changes the
properties of the model drastically, now exhibiting a confining phase.
We finally comment on our findings in Sec. \ref{comment}.

One last word of caution: it is obvious from the very beginning that
the one-loop approximation performed here is hardly appropriate for
dealing with the strongly coupled gauge systems under consideration.
In fact, the results presented below mostly represent an extreme
extrapolation of perturbation theory to extraordinary large values of
the coupling constant $g$ without any reasonable justification.
Nevertheless, besides being interesting in its own right, the model
can serve as a starting point for more involved investigations. E.g.,
the renormalization group flow of the true effective action will
coincide with the perturbative action at large momentum scales; hence,
a detailed knowledge of the perturbative regime will be of use for
checking nonperturbative solutions. Moreover, some of the technical
results of the present calculation such as the form of the gluon
propagator in a fluctuating $A_0$ background will be expedient for
other problems as well.

\section{The Model}
\label{model}

The essence of the model under consideration is determined by the
choice of the background field, which is treated as a classical field
subject to thermal and quantum fluctuations.  At the very
beginning, we confine ourselves to quasi-abelian background fields,
pointing into a fixed direction $n^a$ in color space:
\begin{equation}
\mathsf{A}_\mu:= A_\mu^a T^a =: A_\mu\, n^aT^a,\qquad n^2=1, \label{2}
\end{equation}
where $(T^a)^{bc}\equiv -\I f^{abc}$ denote the hermitean generators
of the gauge group SU($\Nc$) in the adjoint representation. 

Now we are aiming at a derivative expansion in the time-like
component\footnote{The spatially varying $A_0$ field giving rise to
  an electric field appears to conflict with the assumption of
  thermal equilibrium, which is inherent in the Matsubara formalism
  used below; this is because electric fields tend to separate
  (fundamental) color charges, moving the system away from
  equilibrium. However, we adopt the viewpoint that the
  here-considered vacuum model characterizes only a few features of
  the true vacuum; the latter actually includes quark and magnetic
  gluon condensates (and higher cumulants) which altogether are in
  equilibrium. Beyond this, we expect the present approximation to
  hold for sufficiently weak electric fields, keeping the system close
  to equilibrium. Therefore, the expansion in the electric field
  performed below is consistent with (almost) thermal equilibrium.}
of $A_\mu$; such an expansion is usually justified by demanding that
the derivatives be smaller than the characteristic mass scale
of the theory. However, in the present case, there is no initial mass
scale, since the fluctuating particles, gluons and ghosts, are
massless. In fact, it turns out to be impossible to establish a unique
derivative expansion for the (inverse) gluon propagator by a simple
counting of derivatives; this is because a typical expansion generates
terms $\sim \frac{1}{|\partial_i A_0|}$, acting like a mass scale for
the higher derivative terms. Therefore, we propose a different
expansion scheme that is guided by the residual quasi-abelian gauge
symmetry, which still holds for the background field.

The model is further specified by assuming that there are no
magnetic field components in the rest frame of the heat bath; the
latter is characterized by its 4-velocity vector $u^\mu$. Therefore,
there are only two independent (quasi-abelian) gauge invariants:
\begin{eqnarray}
\mathbf{E}^2 &=&\frac{1}{2}F_{\mu\nu} F_{\mu\nu} \equiv 
F_{\mu\alpha}u_\alpha\, F_{\mu\beta} u_\beta \nonumber\\
\bar{A}_u&:=& \frac{1}{\beta} \int\limits_0^\beta d\tau\,
A_u(x^\mu+\tau u^\mu), \qquad A_u:=A_\mu u_\mu. \label{3}
\end{eqnarray}
Here we work in Euclidean finite-temperature space $R^{d-1}\times
S^1$, and $F_{\mu\nu}$ denotes the quasi-abelian field strength of the
background field. In the heat-bath rest frame, we simply have
$u^\mu=(1,\boldsymbol{0})$, so that $A_u\equiv A_0$. The quantity
$\bar{A}_u$ is invariant under quasi-abelian gauge transformations
\cite{gies99a}, since these transformations are restricted to be
periodic in the compactified time direction. (For the complete gauge
group, $\bar{A}_u$ can be modified by a gauge transformation that
differs at $x_0=0$ and $x_0=\beta$ by a center element, e.g.,
$\bar{A}_u\to \frac{2\pi T}{g}-\bar{A}_u$ for SU(2) modulo Weyl
transformations.)

If we now perform a derivative expansion in the electric field
$\mathbf{E}$, we will obtain an effective action of the form $\Gamma
=f(\bar{A}_u)+g(\bar{A}_u)\, E^2 + {\cal O}(E^4, E\partial^2 E)$,
($E\equiv |\mathbf{E}|$, $f,g$ to be determined) for reasons of gauge
invariance. The indicated higher-order terms are at least of fourth
order in $\partial A$ and will be omitted in the following.

Now, the crucial observation is that there exists a unique choice of
gauge for the background field that (i) satisfies the Polyakov gauge
condition $\partial_0 A_0=0$ in order to ensure the correspondence
between $A_0$ and $L$, and (ii) establishes a one-to-one
correspondence between $\mathbf{E}$ and $A_0$, so that an expansion in
$\mathbf{E}$ can be rated as a derivative expansion in $A_0$ (from now
on, we work in the heat-bath rest frame where $A_0\equiv A_u$):
\begin{equation}
A_0(x) = a_0 -(x-x')_i E_i \qquad \Leftrightarrow \qquad E_i
=-\partial_i A_0(x), \label{4}
\end{equation}
where $a_0$ and $E_i$ are considered as constant, and $x'$ is an
arbitrary constant vector which can be set equal to zero. This gauge
can be viewed as a combination of Polyakov and Schwinger-Fock gauge;
the background field considered here lies exactly where the gauge
conditions overlap.  We remark that this is no longer true for
higher derivative terms. The final task is to integrate out the
thermal and quantum fluctuations in the background of the gauge field
\re{4} and expand to second order in $E_i$.

At this point, it is useful to introduce the dimensionless
temperature-rescaled variable
\begin{equation}
c:= \frac{g A_0}{2\pi T}, \qquad c\in [0,1]. \label{5}
\end{equation}
The compactness of $c$ arises from the fact that $A_0$ is a compact
variable in finite-temperature Yang-Mills theories\footnote{This can
  be inferred from a Hamiltonian quantization starting from the Weyl
  gauge $A_0=0$ and generating an $A_0$ field by a time-dependent
  gauge transformation. This observation will furthermore become
  obvious when studying the background field dependence of the gluon
  propagator (cf. Eq.~\re{gluonprop}).}. Then, the resulting effective
action can be represented as a derivative expansion in $c$:
\begin{equation}
\Geff^T[c] =\int d^dx\, \Bigl( V(c,n^a) +W(c,n^a)\,
\partial_i c\partial_i c \Bigr), \label{6}
\end{equation}
where the potential $V$ and the weight function $W$ also depend on the
color space unit vector $n^a$. Higher-order terms $\sim (\partial
c)^4$ are neglected. The superscript $T$ in Eq.~\re{6} signals that
the effective action strongly depends on the presence of a heat
bath. Indeed, $V$ as well as $W$ vanish or reduce to simple constants
as $T\to 0$; this is because the $A_0$ field (or $\bar{A}_u$ in
Eq.~\re{3}) ceases to be an invariant at $T=0$, since it can be gauged
away completely when the time direction is noncompact.

Already at this stage, typical properties of the model become
apparent. First, we observe that if $W(c,n^a)\geq 0$ fluctuations of
the Polyakov loop are suppressed; then, the ground state is solely
determined by the minimum (or minima) of $V(c,n^a)$, which we denote
by $c_V, n^a_V$. This ground state then is (not) confining if it
corresponds to a center (a-)symmetric state, implying $\langle
L\rangle =0$ ($\langle L\rangle\neq 0$).

Fluctuations of the Polyakov loop can only be preferred if $W(c,n^a)$
becomes negative for certain values of $c$ and $n^a$, which we denote
by $c_W$ and $n^a_W$.  Whether or not these fluctuations lead to a
confining phase again depends on the question of whether or not the
minimum of $W(c,n^a)$ corresponds to a center-symmetric state.
Moreover, it depends on the question of whether these fluctuations are
strong enough to compensate for the influence of $V(c,n^a)$. Here we
arrive at a main problem of the model: if $W(c_W, n^a_W)<0$, then the
action \re{6} is not bounded from below. In other words, arbitrarily
strong fluctuations of $c$ around $c_W$ will lower the action without
any bound. Of course, it is reasonable to assume (which we do in the
following) that higher derivative terms $(\partial c)^4$ or
$(\partial^2 c)^2$ will establish such a lower bound, so that the
strength of the fluctuations is dynamically controlled.

Nevertheless, one drawback remains: we cannot make any statement about
the nature of a possible phase transition. For this, we would have to
know everything about the dynamical increase of $\partial_i
c\partial_i c$ when $W(c_W,n^a_W)$ becomes negative for certain values
of temperature.  Since this is beyond the capacities of our model, we
shall always assume in the following that the system is dominated by
the weight function $W$ and thus by fluctuations of the Polyakov loop
whenever $W$ becomes negative.

Let us finally perform a dimensional analysis of the model. For
simplicity, let us start with $d=4$. With regard to Eq.~\re{6}, the
potential has mass dimension $4$, while the weight function has mass
dimension $2$. Due to the compactness of $A_0$ as reflected by Eq.
\re{5}, the only mass scale which is {\em a priori} present is given
by the temperature $T$. Hence, if $V$ scaled with $T^4$ and $W$ with
$T^{2}$, say $V(c,n^a)=T^4\, v(c, n^a)$ and $W(c, n^a)=T^{2}\,
w(c,n^a)$, where $v,w$ are independent of $T$, then we would never
encounter a phase transition in our model; this is because increasing
or lowering the temperature could never turn $W$ from positive
to negative values or vice versa.

At this stage, one may speculate that, since scale invariance is
broken in Yang-Mills theories, the phenomenon of dimensional
transmutation introduces another scale $\mu$ (e.g., the scale at which
the renormalized coupling is defined). Then, the dimensionless
function $w$ can also depend on $T/\mu$. However, this is far from
self-evident, since the breaking of scale invariance is induced by UV
effects. But the functions $V$ and $W$ in the effective action
$\Geff^T$ arise at finite temperature only and thus are a
product of infrared physics. In particular, there are no UV
divergences in the finite-temperature contributions to
$\Geff$ which require another scale during a
regularization procedure. Hence, one is tempted to conclude that the
naive scaling argument given above is correct.

Nevertheless, the naive scaling breaks down, as we shall see in the
next section; but this time, an additional scale is introduced by the
properties of the theory in the infrared. As is well known,
finite-temperature field theories can develop a more singular infrared
behavior than their zero-temperature counterparts. Indeed, while the
effective action at zero temperature and even the effective action for
thermalized purely magnetic background fields do not suffer from
infrared divergences, the case considered here involving thermalized
electric fields exhibits such singularities, which must be handled
carefully. The massless gluon does not provide for a natural infrared
cutoff which could control the low-momentum behavior of the theory.

To conclude, the $d=4$ model is in principle capable of describing a
phase transition, because the finite-temperature infrared divergences
require an additional scale which introduces distinct high- and
low-temperature domains. Of course, there are various ways to deal
with the infrared singularities; and as we will demonstrate below,
they can arise from different physical motivations, leading to
different physical results. Two possibilities are proposed in the
present work. In the first and more natural one, we regularize the
infrared divergences in the same technical way as the ultraviolet
ones, so that in toto there is only one more scale than in the
classical theory, which we identify with the defining scale of the
coupling constant. As a consequence and a consistency check, the
running of the coupling with temperature coincides with the running of
the coupling with field strength or momenta -- they are characterized
by the same $\beta$-function. In the second possibility, we study by
way of example a regularization of the infrared divergences by an
effective gluon mass $\meff$ which we insert by hand, assuming that
such an additional scale may be generated dynamically in the full
theory. The latter version of the theory exhibits the desired
properties of two phases separated by a deconfinement phase
transition, while the former does not.

At $d>4$ the situation is somewhat different and simpler, since here
the coupling constant $g$ is dimensionful, so that two scales are
present already at the classical level. Moreover, no additional scale
will be introduced at the quantum level, because the theory is
infrared finite for $d>4$.

\section{Calculation of the Effective Action}
\label{calculus}

Starting from the standard formulation of Yang-Mills theories via the
functional integral in Euclidean space with compactified time
dimension, we employ the background field method \cite{abbo81} to fix
the gauge for the fluctuating gluon fields, but thereby maintain gauge
invariance for the background field. We arrive at the one-loop
approximation by neglecting cubic and quartic terms in the fluctuating
fields. The remaining two integrals over the gluonic and ghost
fluctuations are Gaussian and lead to functional determinants upon
integration; the one-loop effective action depending on the background
field then reads
\begin{equation}
\Geff^1[A] =\frac{1}{2} \Tr_{x\text{cL}}\, \ln
\Delta^{\text{YM}}[A]^{-1} 
-\Tr_{x\text{c}} \, \ln \Delta^{\text{FP}}[A]^{-1}, \label{7}
\end{equation}
where $\Delta^{\text{YM}}[A]^{-1}$ denotes the inverse gluon
propagator, and $\Delta^{\text{FP}}[A]^{-1}$ the inverse ghost
propagator, i.e., the Faddeev-Popov operator. The traces run over
coordinate ($x$), color (c), and Lorentz (L) labels. Introducing the
abbreviations $D^2:=D_\mu D_\mu$ and $(DD)_{\mu\nu}:=D_\mu D_\nu$,
where the covariant derivative is defined by $D_\mu:=\partial_\mu -\I
g\mathsf{A}_\mu$, and suppressing the indices, the explicit
representations of the propagators read
\begin{eqnarray}
\Delta^{\text{YM}}_{\text{E}}[A]^{-1}&=& -\left[ D^2 -2\I g\, F
  +\left( \frac{1}{\alpha} -1 \right)\, DD \right],\label{8}\\
\Delta^{\text{FP}}_{\text{E}}[A]^{-1}&=& -D^2, \label{9}
\end{eqnarray}
In the following, we will work in the Feynman gauge, $\alpha=1$, which
simplifies the calculations considerably\footnote{For covariant
  constant background fields, the effective action is known to be
  independent of the gauge parameter \cite{hans87}. In the present
  approximation, the gauge field \re{4} is covariant constant, so that
  we are allowed to set $\alpha=1$; this would no longer remain true
  if we were interested in the higher-derivative terms.}. For the
evaluation of Eq.~\re{7}, the spectrum of the inverse propagators is
required. In color space, diagonalization can be achieved by
introducing the eigenvalues $\nu_l$ of the matrix $n^a (T^a)^{bc}$,
$l=1,\dots,\Nc^2-1$.  The basic building block of the operators in
Eqs.~\re{8} and \re{9} is the covariant Laplacian, which upon
insertion of the background field \re{4} yields (we set $x'=0$)
\begin{equation}
(-\I D[A])^2 = (-\I \partial_i)^2 + (-\I D_0[a_0])^2
+2g\nu_l E_i (-\I D_0[a_0]) x_i + (g\nu_l)^2 E_iE_j\,
x_i x_j,\label{10} 
\end{equation}
where $-\I D_0[a_0]=-\I \partial_0 -g\nu_l a_0$, and the roman indices
run over the $d-1$ spatial components. The operator is obviously of
harmonic oscillator type and can be diagonalized by a rigid rotation
of the spatial part of the coordinate system. A prominent eigenvector
is given by the direction of the electric field $E_i$, which we may
choose to point along the $1^{\text{st}}$ direction of the new system.
We finally obtain
\begin{equation}
(-\I D[A])^2 =(-\I \partial_2)^2 +\dots + (-\I \partial_{d-1})^2 +
\bigl( g\nu_l E x_1 +(-\I D_0)\bigr)^2 +( -\I \partial_1)^2,\label{11}
\end{equation}
where $E=\sqrt{E_iE_i}$. Up to now, we have achieved a partial
diagonalization of the operators of Eq.~\re{8} and \re{9}. While the
Faddeev-Popov operator coincides with the Laplacian \re{11}, the
inverse gluon propagator receives additional contributions from the
gluon-spin coupling to the electric field $\sim -\I gF_{\mu\nu}$ which
can easily be diagonalized.  Performing a Fourier transformation for
the $d-2$ unaffected components, $-\I
\partial_2,\dots,-\I\partial_{d-1}\to p_2,\dots, p_{d-1}$, as well as
for the time derivative, $-\I \partial_0 \to \omega_n$, $\omega_n=2\pi
Tn$, $n\in \mathbbm{Z}$ (Matsubara frequencies), we may write the
inverse gluon propagator in the form
\begin{equation}
\Delta^{\text{YM}}[A]^{-1} =p_2^2+\dots+p_{d-1}^2 +\bigl( g\nu_l E\,
x_1 +(\Pi_0)\bigr)^2 +(-\I \partial_1)^2 +2\lambda g\nu_l E,
\label{gluonprop}
\end{equation}
where $\Pi_0=\omega_n -g\nu_l a_0$\footnote{The compactness of $A_0$
  or $a_0$ becomes obvious here; e.g., for SU(2), where
  $\nu_l=-1,0,1$, a shift of $a_0$ by an integer multiple of $(2\pi
  T)/g$ can be compensated for by a shift of the Matsubara label
  $n$.}. The number $\lambda$ labels the different eigenvalues in
Lorentz space arising from the above-mentioned gluon spin coupling
with $\lambda=1,-1,0$; here, $\lambda=1,-1$ appears only once, whereas
$\lambda=0$ occurs with multiplicity $d-2$, corresponding to the
spatial directions which are unaffected by the electric field.
Incidentally, the Faddeev-Popov operator is identical to Eq.
\re{gluonprop} with $\lambda=0$ and multiplicity 1. Taking the
prefactors and signs of the two traces in Eq.~\re{7} into account, the
Faddeev-Popov operator cancels exactly against two Lorentz eigenvalues
of the spectrum of $\Delta^{\text{YM}}[A]^{-1}$ with $\lambda=0$,
removing the spurious gauge degrees of freedom, so that only the
physical, transverse part of the inverse gluon propagator remains,
\begin{equation}
\Delta^{\text{YM}}_\bot[A]^{-1} =p_2^2+\dots+p_{d-1}^2 +\bigl( e_l\,
x_1 +(\Pi_0[a_l])\bigr)^2 +(-\I \partial_1)^2 +2\lambda e_l,
\label{13}
\end{equation}
where $\lambda=0$ now occurs with multiplicity $d-4$. For reasons
of brevity, we introduced the short forms
\begin{equation}
e_l :=|g\nu_l E|, \qquad a_l:= |g\nu_l a_0| \label{14}
\end{equation}
in Eq.~\re{13}; the use of the moduli in Eq.~\re{14} is justified by
the observation that, when tracing over a function of the inverse
propagators, the result will not be sensitive to the signs of $g\nu_l
E$ and $g\nu_l a_0$.

The remaining problem of diagonalizing the $0$-$1$ subspace at first
sight resembles the problem of finding the spectrum of a relativistic
particle in a constant magnetic field.  There, one finds the
eigenvalues (Landau levels) by shifting the $x_1$ coordinate by
$x_1\to x_1 -\frac{(-\I \Pi_0)}{e_l}$ in order to arrive at a
perfect harmonic oscillator. Here, the situation is not so simple,
because the $a_0$ field as well as the temperature dependence would
drop out of the operator completely. In other words, such a shift is
not in agreement with the periodic boundary conditions in time
direction.

Hence, the usual harmonic oscillator techniques arrive at their limits,
and we have to find a different method that does not rely on the
explicit knowledge of the spectra as is necessary for, e.g.,
$\zeta$-function methods. We choose Schwinger's proper-time technique,
which provides for a more direct handling of the propagators. In terms
of the transverse gluon propagator, the effective action reads in
proper-time representation
\begin{eqnarray}
\Geff^1[A]&=& \frac{1}{2} \Tr_{x\text{cL}}\, \ln \Delta^{\text{YM}}_\bot
[A]^{-1}=-\frac{1}{2} \tr_{\text{cL}} \int\limits_0^\infty
\frac{ds}{s}\, \langle x| \E^{-s\Delta^{\text{YM}}_\bot
[A]^{-1}}|x\rangle \nonumber\\
&\equiv& -\frac{1}{2} \tr_{\text{cL}} \int\limits_0^\infty
\frac{ds}{s} \, \Omega \int\!\!\!\!\!\!\!\!\sum
 \frac{d^d p}{(2\pi)^d}\, \E^{-s M(p,\lambda,l;s)},
\label{15}
\end{eqnarray}
where $\Omega$ denotes the spacetime volume of $\mathbbm{R}^{d-1}
\times S^1$, $s$ is the proper time, and the trace over the continuous
part of the spectrum is taken in momentum space. The color trace runs
over $l$, which labels the color space eigenvalues, whereas the Lorentz
trace runs over $\lambda$ with its associated multiplicities. The
function $M$ is defined via the Fourier representation of the
proper-time transition amplitude
\begin{equation}
\langle x| \E^{-s\Delta^{\text{YM}}_\bot[A]^{-1}}|x'\rangle
=\int\!\!\!\!\!\!\!\!\sum \frac{d^d p}{(2\pi)^d} \, \E^{\I p(x-x')}\,
\E^{-s M(p,\lambda,l;s)}, \label{16}
\end{equation}
which can be determined by the differential equation
\begin{equation}
\mathbbm{1}= \Delta^{\text{YM}}_\bot [A]^{-1} \,
\Delta^{\text{YM}}_\bot [A]. \label{17}
\end{equation}
When evaluated, for example, in momentum space, Eq.~\re{17} is solved
by 
\begin{equation}
\Delta^{\text{YM}}_\bot [A](p,\lambda,l) =\int\limits_0^\infty ds\,
\E^{-s M(p,\lambda,l;s)}, \label{18}
\end{equation}
where $M$ is given by
\begin{eqnarray}
M(p,\lambda,l;s) &=& p_2^2 +\dots+p_{d-1}^2 +\frac{\tanh 2e_l s}{2e_ls}
(p_1 +q)^2 + \frac{\tanh e_l s}{e_ls} (\omega_n -a_l)^2 \nonumber\\
&&+ \frac{1}{2s} \ln \cosh 2e_l s +2e_l \lambda. \label{19}
\end{eqnarray}
Here, $q$ denotes some function of $e_l$ and $s$ which becomes
irrelevant when shifting the $p_1$ integration in Eq.~\re{15}. Upon
insertion of Eq.~\re{19} into Eq.~\re{15}, the Gaussian momentum
integration and the sum over $\lambda$ can easily be performed; the
sum over Matsubara frequencies can be reorganized by a simple Poisson
resummation,\footnote{For technical details, see, e.g.,
  \cite{gies99a,DG00}.} and we arrive at
\begin{eqnarray}
\Geff^1[A]&=& -\frac{\Omega}{2} \tr_{\text{c}} \frac{1}{(4\pi)^{d/2}}
\int\limits_0^\infty \frac{ds}{s^{d/2}}\, e_l\left( 4\sinh e_l s +
  \frac{d-2}{\sinh e_l s} \right) \nonumber\\
&&\qquad\qquad\qquad\quad \times \left[ 1+2 \sum_{n=1}^\infty \exp
  \left( -\frac{n^2}{4T^2} e_l \coth e_l s\right) \cos \frac{a_l}{T} n 
\right]. \label{20}
\end{eqnarray}
Here, we have separated the zero-temperature part, corresponding to the
first line times the ``1'' of the second line, from the
finite-temperature contributions, corresponding to the first line read
together with the $n$ sum.  

\subsection{Effective Action at Zero Temperature}
Let us first study the temperature-independent part of the effective
action Eq.~\re{20} with particular emphasis on its renormalization:
\begin{equation}
\Geff^{1T=0}[A]= -\frac{\Omega}{2} \tr_{\text{c}} \frac{1}{(4\pi)^{d/2}}
\int\limits_0^\infty \frac{ds}{s^{d/2}}\, e_l\left( 4\sinh e_l s +
  \frac{d-2}{\sinh e_l s} \right). \label{21}
\end{equation}
On the one hand, the proper-time integral is divergent at the upper
bound, $s\to \infty$, owing to the first term $\sim \sinh e_l s$.
Since large values of $s$ correspond to the infrared regime, this
divergence is not related to the standard renormalization of bare
parameters, which is a UV effect. In fact, this divergence is
analogous to the Nielsen-Olesen unstable mode \cite{niel78} of the
Savvidy vacuum\footnote{At $T=0$, the present situation involving an
  external {\em electric} field is identical to the {\em magnetic}
  Savvidy vacuum owing to the Euclidean $O(4)$ symmetry.}; one can
give a meaning to this essential singularity by rotating the contour
of the integral over the $\sinh$ term into the lower complex plane,
$-\I s\to s$. The effective action then picks up an imaginary part
that characterizes the instability of the constant electric background
field considered here.

On the other hand, the proper-time integral is also divergent at the
lower bound, corresponding to the ultraviolet. The leading singularity
is of the order $s^{-d/2}$, so that $m$ subtractions are required for
$d=2m$ or $d=2m+1$. The leading singularity which is field independent
can easily be removed by demanding that $\Geff^{T=0}[A=0]=0$ (first
renormalization condition). The next-to-leading singularity
proportional to $e_l^2\sim E^2$ is removed by the second
renormalization condition $(\partial \Leff/\partial E^2)|_{E\to
  0}=1/2$, where $\Geff=\int \Leff$; this ensures that the classical
Lagrangian is recovered when all nonlinear interactions are switched
off, and corresponds to a field-strength and charge renormalization
\begin{equation}
{\cal L}_{\text{cl}}^{\text{R}} \equiv \frac{1}{2} E_{\text{R}}^2
=\frac{1}{2} Z_3^{-1} E^2, \label{22}
\end{equation}
where $E_{\text{R}}$ denotes the renormalized field, and $Z_3$ is the
wave function renormalization constant. The latter can be read off
from Eq.~\re{21} by isolating the singularity $\sim E^2$,
\begin{eqnarray}
Z_3^{-1}&=& 1-\frac{26 -d}{6(4\pi)^{d/2}} \, \Nc\, \bar{g}^2
\int\limits_{\mu^2/\Lambda^2} \frac{ds}{s^{d/2-1}}, \label{23}
\end{eqnarray}
where we have used an explicit cutoff $\Lambda$, employed
$\tr_{\text{c}} |\nu_l|^2 =\sum_{l=1}^{\Nc^2-1} |\nu_l|^2 =\Nc$, and
have introduced the dimensionless coupling $\bar{g}^2= g^2 \mu^{d-4}$
with the aid of a reference scale $\mu$ (at which $g$ is defined). To
one-loop order, the $\beta$ function can be read off from the
coefficient of the UV divergence of $Z_3^{-1}$:
\begin{equation}
\beta_{\bar{g}^2}\equiv\partial_t \bar{g}^2 =(d-4) \bar{g}^2 -b_0^d
\bar{g}^4,\label{24}
\end{equation}
where
\begin{eqnarray}
b_0&=&\frac{11}{3} \frac{\Nc}{8\pi^2}, \quad \text{for}\quad
d=4,\nonumber\\
b_0^d&=&\frac{(26-d)}{3(d-4)} \frac{\Nc}{(4\pi)^{d/2}}, \quad
\text{for}\quad d>4,\label{25} 
\end{eqnarray}
and $t$ denotes the ``renormalization group time'' $\ln \mu/\Lambda$.
Here we have rediscovered the standard well-known one-loop results,
including the remarkable observation that the $\beta$ function for the
dimensionful coupling $g^2=\bar{g}^2 \mu^{4-d}$ vanishes precisely in
the critical string dimension $d=26$ \cite{frad83}. 

Note that in $4<d<26$, the $\beta$ function develops a UV-stable
fixed point:
\begin{equation}
\bar{g}_\ast^2 =\frac{d-4}{b_0^d}=\frac{3(d-4)^2}{26-d} \,
\frac{(4\pi)^{d/2}}{\Nc}. \label{26}
\end{equation}
Of course, this fixed point lies in the perturbative domain
($\bar{g}_\ast^2/4\pi\ll 1$) only for very large $\Nc$.

As an alternative to these considerations of renormalization,
the integral in Eq.~\re{21} can be treated more directly with an
appropriate regularization prescription. Let us briefly sketch a
proper-time variant of dimensional regularization for later use in the
case $d=4$. Shifting the singularities at $s\to 0$ by $\epsilon$ and
introducing a mass scale $\mu$, Eq.~\re{21} can be written as
\begin{equation}
\Leff^{1T=0}= -\frac{1}{8\pi^2} \tr_{\text{c}}\, \mu^{2\epsilon}\left[
  \int\limits_0^{-\I\infty} \frac{ds}{s^{2-\epsilon}}\, e_l \sinh e_l
  s + \frac{d-2}{4} \int\limits_0^\infty \frac{ds}{s^{2-\epsilon}}\,
  \frac{e_l}{\sinh e_l s} \right]. \label{27}
\end{equation}
These integrals can be evaluated \cite{GR}, and the result for the
one-loop contribution to the zero-temperature effective Lagrangian in
$d=4$ reads
\begin{equation}
\Leff^{1T=0}=-\frac{1}{8\pi^2} \tr_{\text{c}} \, e_l^2 \left[
  \frac{11}{12\epsilon} -\frac{11}{12} \ln \frac{e_l}{\mu^2}
  +\text{const.} +\text{imag. parts} +{\cal
  O}(\epsilon)\right]. \label{28}
\end{equation}
The appearance of the simple pole in $\epsilon$ implies a charge and
field strength renormalization as outlined above. To be precise, in
the background field formulation, the coupling runs with the scale set
by the strength of the external {\em field}: $g^2=g^2(gE/\mu^2)$; this
is analogous to the {\em momentum} dependence of the coupling in the
standard formulation.  Including the correctly (re-)normalized
classical term, the total effective Lagrangian to one loop can then be
written as
\begin{equation}
\Leff^{T=0}(gE)={\cal L}_{\text{cl}}+\Leff^{1T=0} =\frac{1}{8} b_0\,
(gE)^2 \ln \frac{(gE)^2}{\E \kappa^2}+\text{imag. parts}, \qquad
\kappa^2 =\mu^4 \E^{-\frac{4}{b_0g^2} -1}, \label{29}
\end{equation}
where we have introduced the renormalization group invariant quantity
$\kappa$ corresponding to the minimum of $\Leff^{T=0}(gE)$, and $b_0$
is given by the first line of Eq.~\re{25}. 

Concerning the imaginary parts, the following comment should be made:
within the Savvidy model, the imaginary parts indicate the instability
of the constant-field vacuum configuration signaling the final failure
of the model. In the present case, they are just an artefact of
truncating the derivative expansion of the effective action at second
order; this truncation is formally equivalent to the constant-field
approximation. Upon an inclusion of non-constant terms which would
affect only higher-derivative contributions, we expect the imaginary
part to vanish; this is because the unstable modes are then cut off by
the length scale of variation of the fields.

For the regularization/renormalization program of $\Leff^{1T=0}$ in
$d\geq6$, more subtractions than in $d=4$ are needed; since these
theories are nonrenormalizable in the common sense, these subtractions
correspond to counter-terms of field operators of higher mass
dimension. To be precise, $d>4$ Yang-Mills theories can only be
defined with a cutoff (with physical relevance); therefore, some
cutoff procedure is implicitly understood. Nevertheless, the precise
form of the cutoff procedure only affects higher-order operators which are
of no relevance for the present model. Moreover, it is perfectly
legitimate to study $d>4$ quantum Yang-Mills theories in the sense of
effective theories valid below a certain cutoff scale.

\subsection{Effective Action at Finite Temperature}
Equipped with these preliminaries, we now turn to the more interesting
finite-temperature part of Eq.~\re{20}:
\begin{eqnarray}
\Leff^{1T}&=&-\frac{1}{(4\pi)^{d/2}} \tr_{\text{c}} 
\int\limits_0^\infty \frac{ds}{s^{d/2}}\, e_l\left( 4\sinh e_l s +
  \frac{d-2}{\sinh e_l s} \right) 
\sum_{n=1}^\infty \exp
  \left( -\frac{n^2}{4T^2} e_l \coth e_l s\right) \cos \frac{a_l}{T}
  n. \nonumber\\
  \label{30}
\end{eqnarray}
For $s\to 0$, the integral remains completely finite, since the
$\coth$ in the exponent develops a $1/s$ pole; i.e., there are no UV
divergences in the thermal contribution to $\Leff$, as is to be
expected. 

At the opposite end, $s\to\infty$, we again encounter the $\sinh$
divergence induced by the unstable mode. However, this is not the only
infrared problem: an attempt at circumventing this problem by a rotation
of the $s$ contour as in the zero-temperature case would lead to a
disastrous behavior of the $n$ sum due to the poles of the $\coth$ on
the imaginary axis. In fact, it is the interplay between the
proper-time integral and the $n$ sum that produces further infrared
divergences (at least for d=4). It is well known in the literature
that particles with Bose-Einstein statistics develop stronger infrared
singularities at $T\neq0$ than at zero temperature
\cite{lebe96}. Unfortunately, the status of these finite-temperature
singularities is far from being settled, contrary to the $T=0$ case. 

In the present paper, we shall investigate two different methods. The
consequences of an explicit mass-like cutoff are discussed in Sec.
\ref{gluonmass}. Here, we propose a more natural treatment by
regularizing the thermal infrared divergences of Eq. \re{30} by the
same method used to treat the UV divergences in Eq. \re{27} in the
$T=0$ case. Thereby, the same scale $\mu$ which serves to define the
value of the coupling constant is introduced.

Taking these considerations into account, the Lagrangian is modified
according to (substitution $\mu^2 s= u$)
\begin{eqnarray}
\Leff^{1T}&=&-\frac{4}{(4\pi)^{d/2}} \tr_{\text{c}} \, \mu^d
\int\limits_0^\infty \frac{du}{u^{d/2-\epsilon}} \left(
  \frac{e_l}{\mu^2} \sinh \frac{e_l}{\mu^2} u+ \frac{d-2}{4}
  \frac{e_l/\mu^2}{\sinh \frac{e_l}{\mu^2} u} \right)\nonumber\\
&&\qquad\qquad\qquad\quad\times 
\sum_{n=1}^\infty \exp\left( -\frac{n^2}{4} \frac{\mu^2}{T^2}
\frac{e_l}{\mu^2} \coth \frac{e_l}{\mu^2} u \right) \cos
\frac{a_l}{T}n.
\label{31}
\end{eqnarray}
In the context of our approximation in terms of derivatives of $A_0$,
we need only the terms $\sim e_l^0$ and $\sim e_l^2$ of
Eq.~\re{31}. Expanding in $e_l/\mu^2$ and performing the $s$
integral, we arrive at
\begin{eqnarray}
\Leff^{1T}\bigr|_0&=&-\frac{(d-2)\Gamma(d/2)}{\pi^{d/2}}
\sum_{l=1}^{\Nc^2-1} \sum_{n=1}^\infty \frac{\cos \frac{a_l}{T} n}{n^d}
\,T^d =: V(c,n^a), \label{32}\\
\Leff^{1T}\bigr|_{e_l^2}&=& -\frac{1}{6\pi^{d/2}} 
\sum_{l=1}^{\Nc^2-1}  \left(\frac{e_l}{\mu^2}\right)^2 
\sum_{n=1}^\infty \frac{\cos \frac{a_l}{T} n}{n^d}
\left(\frac{n^2\mu^2}{4T^2}\right)^{2+\epsilon} \nonumber\\
&&\qquad\qquad\times\Gamma(d/2 \!-\!2
\!-\!\epsilon) \bigl[ (26-d) -(d-2)(d-4-2\epsilon)\bigr] T^d,
\label{33}
\end{eqnarray}
For the term $\sim e_l^0$ in the first line, the $\epsilon\to0$ limit
could safely be performed for $d\geq0$; by construction, this term
depends only on $a_l\sim a_0\sim c$ (cf. Eq.~\re{5}) and therefore
corresponds to the potential $V(c,n^a)$ as introduced in Eq.~\re{6}. 

The term $\sim e_l^2$ in the second line contributes to the function
$W(c,n^a)$ (in addition to the classical term). It turns out that, for
$d>4$, the limit $\epsilon\to 0$ can be performed immediately without
running into an $\epsilon$ pole. This means that, in these dimensions,
the thermally modified infrared behavior of the theory is under
control. The order $e_l^2$ term of the one-loop effective action then
reads
\begin{equation}
\Leff^{1T}\bigr|_{e_l^2} =-\frac{\Gamma(d/2-2)}{96 \pi^{d/2}}
\sum_{l=0}^{\Nc^2-1} \left( \frac{e_l}{T^2}\right)^2 \sum_{n=1}^\infty
\frac{\cos \frac{a_l}{T} n}{n^{d-4}} \bigl[ (26\!-\!d)
-(d\!-\!2)(d\!-\!4)\bigr] T^d, \quad d>4. \label{34}
\end{equation}
Obviously, the $\mu$ dependence has dropped out as a consequence
of the well-behaved $\epsilon\to 0$ limit. Nevertheless, there is a
second scale besides the temperature, which is given by the
dimensionful coupling constant $g$ in $d>4$.

In $d=4$, the situation is more involved, since Eq.~\re{33} develops
a simple pole in $\epsilon$ for $\epsilon\to 0$. In order to isolate
the pole and the terms of order $\epsilon^0$ which contain the
physics, we first have to perform the $n$ sum; this can be achieved
with the aid of the polylogarithmic function (also Jongqui\`{e}res
function) 
\begin{equation}
\Li(z,q):=\sum_{n=1}^\infty \frac{q^n}{n^z} \label{35}
\end{equation} 
and its analytical continuation for arbitrary real values of $z$
\cite{math}. We finally find for Eq.~\re{33} in $d=4$:
\begin{eqnarray}
\Leff^{1T}\bigr|_{e_l^2}&=& -\frac{1}{8\pi^2} \sum_{l=1}^{\Nc^2-1} e_l^2
\left[ \frac{11}{12\epsilon} - \frac{11}{12} \ln \frac{T^2}{\mu^2} +
  \frac{11}{6} \Li'(0,\E^{\I \frac{a_l}{T}}) +\frac{11}{6}
  \Li'(0,\E^{-\I \frac{a_l}{T}})  \right.\nonumber\\
&&\qquad\qquad\qquad\quad \left.+\frac{1}{6} + \frac{11}{12} C
  -\frac{11}{12} \ln4 
\right],\quad d=4, \label{36}
\end{eqnarray}
where the prime at $\Li$ denotes the derivative with respect to the
first argument, and $C$ is Euler's constant $C\simeq0.577216$. Our
first observation is that the $\epsilon$ pole in this thermal
contribution is identical to the one for the zero-temperature
Lagrangian in Eq.~\re{28}. Since the latter is responsible for the
usual charge and field strength renormalization leading to a
field-strength-dependent coupling $g^2=g^2(gE/\mu^2)$, the present
$\epsilon$ pole analogously suggests a running of the coupling with
the scale set by the temperature: $g^2=g^2(T^2/\mu^2)$. And because
the residues of each pole are identical, the thermal running is
governed by the same $\beta$ function. This can be viewed as a
consistency check of our treatment of the infrared singularities.

Furthermore, the terms $\sim \epsilon^0$ depend on the ratio
$T^2/\mu^2$ (even in the limit $a_l\to 0$). This implies that they
cannot be normalized away as in the zero-temperature case, but
lead to a thermal renormalization of the two-point function. 

This is in perfect analogy to QED, where an equivalent modification of
the two-point function appears with the prefactor (= Yang-Mills
$\beta$ function) replaced by the QED $\beta$ function, and the role
of $\mu$ is played by the natural scale of QED: the electron mass
\cite{elmf98,DG00}.

In conclusion, it is the $\ln \frac{T^2}{\mu^2}$ term in Eq.~\re{36}
which leads to a breakdown of the naive scaling as outlined in Sec.
\ref{model} and allows for a separation of high- and low- temperature
regimes. This could in principle facilitate a description of a phase
transition within the $d=4$ model. However, as we shall find in the
next section, the model does not make use of this option.

\section{Analysis of the Effective Action}
\label{analysis}

In the following analysis of the previously derived effective action
for arbitrary $d$ and $\Nc$, for simplicity we confine ourselves to
$\Nc=2$, which provides for a convenient study of all the essential
features of the model. Then, the color space eigenvalues $\nu_l$ are
simply given by
\begin{equation}
\nu_l=-1,0,1,\quad \text{for}\quad \text{SU}(2).
\end{equation} 

The results given above can be summarized in the effective Lagrangian
(cf.  Eqs.~\re{5} and \re{6}):
\begin{equation}
\Leff^T[c] =  V(c) +W(c)\,
\partial_i c\partial_i c,
\end{equation}
where we have used the relations (cf. Eq.~also \re{4})
\begin{equation}
c=\frac{ga_0}{2\pi T}, \quad\text{and}\quad 
\partial_i c=\frac{-gE_i}{2\pi T}, \quad c\in [0,1]. \label{37}
\end{equation}
The convenient dimensionless quantity $c$ is now considered as the
dynamical variable of the effective theory; for SU(2), the center
symmetric point is given by $c=1/2$, since center symmetry relates $c$
with $1-c$. If the vacuum state is characterized by $c=1/2$, our model
is confining, whereas a vacuum state different from $c=1/2$
characterizes the deconfinement phase. 

\subsection{Four Dimensions $d=4$}
\label{4.1}

\begin{figure}
\begin{picture}(145,50)
\put(-3,0){
\epsfig{figure=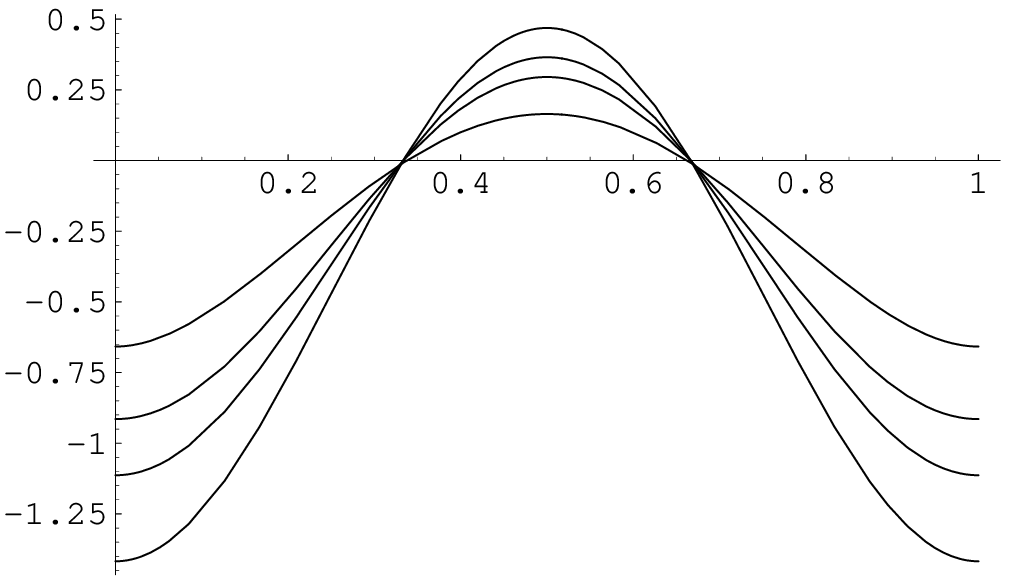,width=7.8cm}}
\put(0,50){(a)}
\put(9,45){$V(c)$}
\put(74,37){$c$}
\put(70,24){$d=$}
\put(75,19){4}
\put(75,13){7}
\put(75,8){8}
\put(75,2){9}
\put(80,0){
\epsfig{figure=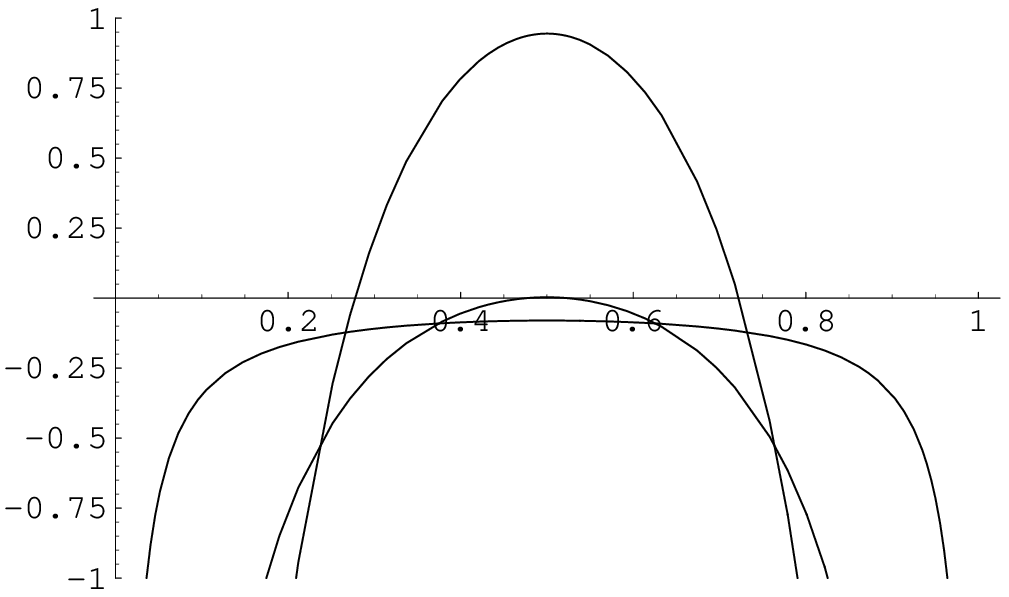,width=7.8cm}}
\put(82,50){(b)}
\put(92,45){$W(c)$}
\put(157,26){$c$}
\put(128,45){$t=1$}
\put(117,26){$t=0.6$}
\put(117,18){$t=0.2$}
\end{picture}
\caption{(a) SU(2)-Polyakov loop potential $V(c)$ in units of $T^d$ for
  $d=4,7,8,9$ (cf. Eqs.~\re{weiss} and \re{42}). The $d=4,8$
  potentials correspond to Bernoulli polynomials $B_4$ and $B_8$. \qquad
  $\,\,\,\,\,$(b) SU(2) weight function $W(c)$ in units of $\kappa$
  in $d=4$ for 
  different values of the temperature $t:=T/\sqrt{\kappa}=0.2,0.6,1$
  (cf. Eq.~\re{40}). The disconnected absolute maxima at $W(c=0,1)$
  are not depicted.}
\label{figV}
\end{figure}

Beginning with the most relevant case of four spacetime dimensions,
the potential can be read off from Eq.~\re{32}. Performing the $n$ sum
leads to a Bernoulli polynomial,
\begin{equation}
V(c)=-\frac{3\pi^2}{45}\, T^4 + \frac{4\pi^2}{3}\, T^4\,
c^2(1-c)^2, \label{weiss}
\end{equation}
in agreement with \cite{weis81}. While the first term is simply the
free energy of $\Nc^2-1=3$ free gluons, the second models the shape of
the potential revealing a maximum at $c=1/2$ and minima at $c=0,1$ and
thereby characterizing the deconfinement phase (see Fig.
\ref{figV}(a)). However, even if the potential had displayed a minimum
at $c=1/2$, it would have been of no use, since the potential by
itself would remain confining for arbitrarily high temperatures. There
would be no comparative scale separating two different phases. A
Polyakov loop potential depending on $c$ and $T$ only can never model
the deconfinement phase transition of Yang-Mills theories!

The weight function $W(c)$ can be read off from Eq.~\re{36} in
combination with the classical Lagrangian ${\cal
  L}_{\text{cl}}=E^2/2=\frac{2\pi^2T^2}{g^2(\mu)} \partial_i
c\partial_i c$:
\begin{eqnarray}
W(c)\!\!&=&\!\! 2\pi^2T^2 \left\{\! \frac{1}{g^2(\mu)} -b_0\!\left[ -\ln
    \frac{T}{\mu} +\frac{C}{2} +\frac{1}{11} -\ln 2
    +\Li'(0,\E^{2\pi\I c}) +\Li'(0,\E^{-2\pi\I c})\right]\!\right\}
    \nonumber\\
&=&\!\!2\pi^2T^2 b_0\left[ \ln \frac{T}{\sqrt{\kappa}} - \frac{1}{4}
  -\frac{1}{11} - \frac{C}{2} +\ln 2 -\Li'(0,\E^{2\pi\I c})
  -\Li'(0,\E^{-2\pi\I c}) \right], \label{40}
\end{eqnarray}
where $b_0$ denotes the $\beta$ function coefficient given in the
first line of Eq.~\re{25} (for $\Nc=2$). In the second line, we have
expressed the running coupling and the scale $\mu$ by the
renormalization group invariant $\kappa$ defined in Eq.~\re{29}, so
that $W(c)$ is itself renormalization group invariant! In fact,
lowering the temperature can turn the weight function negative for any
value of $c$ so that fluctuations of the Polyakov loop are
energetically preferred for low $T$. However, the confining value
$c=1/2$ always represents a local maximum of the weight function
$W(c)$, as is visible in Fig. \ref{figV}(b). For $c\to0,1$, the weight
function diverges to $-\infty$, but at $c=1,0$ it jumps to its
absolute maxima. Analytically, one finds
\begin{equation}
W\bigl([c=0,1;c=1/2]\bigr) =2\pi^2 T^2 b_0 \left( \bigl[ \ln 4\pi; \ln
  \pi\bigr] -\frac{15}{44} -\frac{C}{2} +\ln \frac{T}{\sqrt{\kappa}}
  \right). \label{41}
\end{equation}
To conclude, although our model indicates that fluctuations of the
Polyakov loop become important at low temperatures, they do not
fluctuate around the confining minimum, but energetically prefer a
center asymmetric ground state for $c$. Hence, our model is not
capable of finding a confinement phase\footnote{The discontinuous
  behavior of the weight function for $c\to0,1$ gives rise to
  speculations. Physically, such behavior is not acceptable (nor
  interpretable); rather, one may expect that some mechanism
  will lead to a wash-out of these singularities unveiling a smooth
  functional form of $W(c)$ for $c\in [0,1]$ (although the origin of
  such a mechanism is still unclear to us). Probably, this will
  lead to a weight function of mexican-hat type with deconfining minima.
  However, with even more reservations, one might speculate upon the
  possibility of a smooth curve for $W(c)$ which directly interpolates
  between the extremal values at $c=0,1/2,1$ given in Eq.~\re{41} with
  a confining minimum at $c=1/2$. Then, the model would exhibit a
  confining phase for small enough temperatures when $W(c)$ becomes
  negative for $c=1/2$. The reason for mentioning such vague
  speculations is to demonstrate how possible predictions could in
  principle arise from the model: following the reasoning of Sec.
  \ref{model}, the temperature of the phase transition is then given by
  $W(c=1/2)|_{T=T_{\text{cr}}}=0$. From Eq.~\re{41}, we obtain:
  $T_{\text{cr}}/\sqrt{\kappa}\simeq 0.60$. Identifying $\kappa$ with
  the string tension $\sigma$ (as it is the case in the leading-log
  model \cite{adle82}), our speculative estimate is in remarkably good
  agreement with the lattice value \cite{latt2},
  $T_{\text{cr}}/\sqrt{\sigma}\simeq 0.69$.}.
Nevertheless, it should be stressed that the present treatment of the
infrared modes is part of the definition of the model, although we
have tried to formulate the present version as ``universally'' as
possible. In fact, the regularization method considered here, which
belongs to the standard class of regularization techniques, guarantees
scheme-independent results. But it is also possible that the infrared
modes are screened by a physical mechanism which involves another
scale and thereby introduces ``nonuniversal'' information. Such
a version of the model is discussed by way of example in
Sec. \ref{gluonmass}. 

\subsection{Beyond Four Dimensions $d>4$}
In spacetime dimensions larger than four, the situation simplifies
owing to the absence of infrared problems. The Polyakov loop potential
is again given by Eq.~\re{32}, which, for $\Nc=2$, reads
\begin{equation}
V(c)=-\frac{(d-2)\Gamma(d/2) \zeta(d)}{\pi^{d/2}}\, T^d
-\frac{2}{\pi^{d/2}} (d-2) \Gamma(d/2) \sum_{n=1}^\infty \frac{\cos
  2\pi c n}{n^d}\, T^d, \label{42}
\end{equation}
where $\zeta(d)$ denotes Riemann's $\zeta$ function. Equation \re{42}
is in perfect agreement with \cite{acto83}, where it is demonstrated
that a representation of $V(c)$ in terms of Bernoulli polynomials of
$d$th degree exists in $d=2,4,6,8,\dots$. We could as well choose
a representation in terms of polylogarithmic functions which
interpolate smoothly between the Bernoulli polynomials. In toto, the
qualitative behavior of $V(c)$ does not change significantly for
different $d$: $V(c=1/2)$ is always a (deconfining) maximum
(cf. Fig. \ref{figV}(a)). 

\begin{figure}
\begin{picture}(145,50)
\put(-3,0){
\epsfig{figure=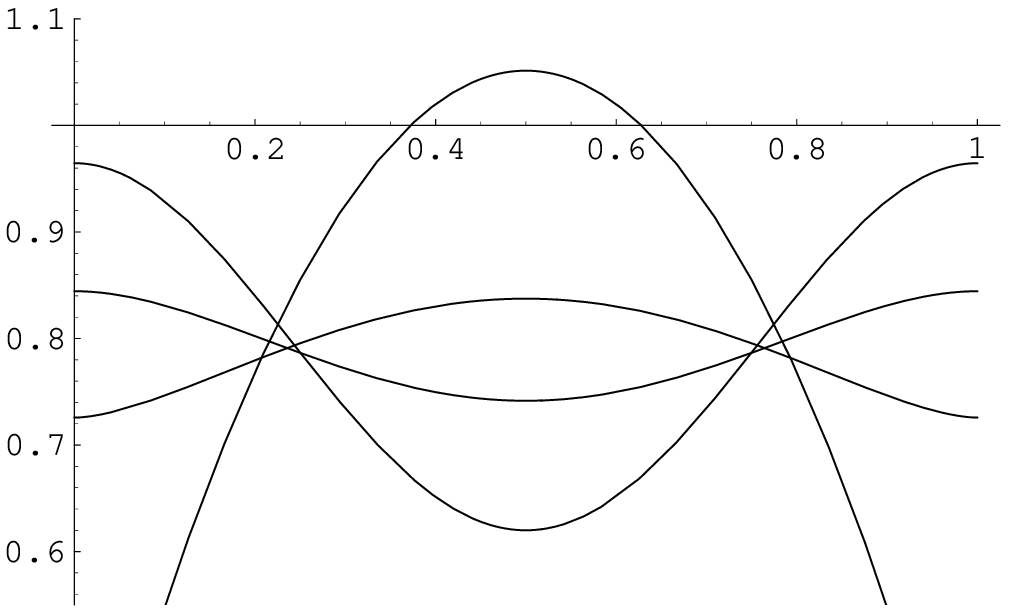,width=7.8cm}}
\put(0,50){(a)}
\put(6,45){$W(c)$}
\put(74,40){$c$}
\put(34,44){$d=6$}
\put(34,26){$d=7$}
\put(34,13){$d=8$}
\put(34,3){$d=10$}
\put(80,0){
\epsfig{figure=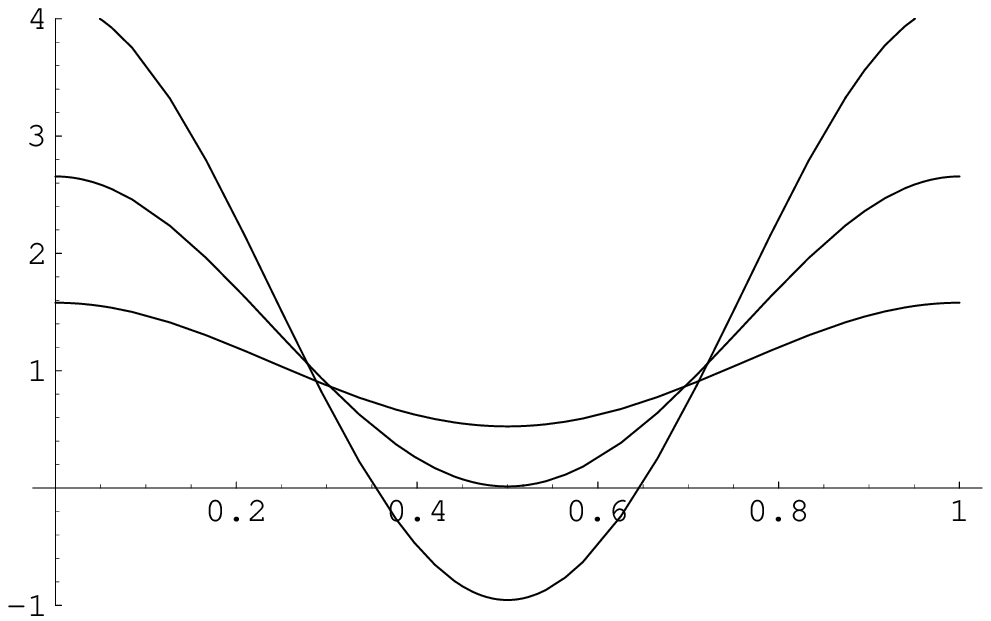,width=7.8cm}}
\put(82,50){(b)}
\put(92,47){$W(c)$}
\put(157,12){$c$}
\put(150,49){$T>T_{\text{cr}}$}
\put(150,36){$T=T_{\text{cr}}$}
\put(150,27){$T<T_{\text{cr}}$}
\end{picture}
\caption{(a) SU(2) weight function $W(c)$ in units of $\mu$ for
  $d=6,7,8,10$ and fixed $T$ and $\bar{g}$ (cf. Eq.~\re{43}). Above
  $d=d_{\text{cr}}\simeq7.42$, $c=1/2$ represents the minimum of
  $W(c)$.\qquad\qquad
  (b) The same weight function is now plotted for fixed
  $\bar{g}$ and $d=10>d_{\text{cr}}$ for various temperature values
  close to $T_{\text{cr}}$.}
\label{figWd}
\end{figure}

The situation is different for the weight function $W(c)$: in terms of
the dimensionless coupling $\bar{g}^2=\mu^{d-4} g^2$ and
polylogarithmic functions, the contributions from Eq.~\re{34}
together with the classical Lagrangian can be represented as
\begin{eqnarray}
W(c)&=& 2\pi^2 \frac{T^2}{\mu^2} \, \mu^{d-2} \Biggl\{
  \frac{1}{\bar{g}^2} - \frac{T^{d-4}}{\mu^{d-4}} \,
  \frac{\Gamma(d/2-2)}{48 \pi^{d/2}} \bigl[ (26\!-\!d)
  -(d\!-\!2)(d\!-\!4) \bigr]  \nonumber\\
&&\qquad\qquad\qquad\qquad\qquad \times \Bigl[ \Li(d-4,\E^{2\pi\I c})
  +\Li(d-4,\E^{-2\pi\I c}) \Bigr]\Biggr\}. \label{43}
\end{eqnarray}
On the one hand, we again encounter the combination of polylogarithmic
functions that interpolate between the Bernoulli polynomials of
$(d-4)$th degree for $d=6,8,\dots$, essentially maintaining their
typical shape. On the other hand, there is an important sign change
owing to the factor $(26\!-\!d) -(d\!-\!2)(d\!-\!4)$ at the ``critical
dimension''
\begin{equation}
d_{\text{cr}}=\frac{1}{2} (5+\sqrt{97})\simeq 7.42. \label{44}
\end{equation}
For $d<d_{\text{cr}}$, $W(c)$ has a maximum at $c=1/2$, implying that
there is no confining phase in these dimensions. But for
$d>d_{\text{cr}}$, the weight function exhibits an absolute minimum at
the center symmetric value $c=1/2$ (see Fig. \ref{figWd}(a)). As a
consequence, $W(c)$ can become negative at $c=1/2$ for {\em increasing}
temperature, as is depicted in Fig. \ref{figWd}(b). This is in agreement
with the fact that the dimensionless coupling grows large in the {\em
  high}-momentum limit with a UV-stable fixed point given by Eq.
\re{26}. Therefore, the model, somewhat counter-intuitively, describes
a system with two different phases, a deconfined phase at low
temperature and a confining strong-coupling phase at high temperature.
In terms of the dimensionful coupling constant, the critical
temperature where $W(c=1/2)|_{T=T_{\text{cr}}}=0$ is given by
\begin{equation}
g^2 T_{\text{cr}}^{d-4}=\frac{24 \pi^{d/2}}{\Gamma(d/2-2) \zeta(d-4)}
\,
\frac{2^{d-5}}{(2^{d-5}-1)\bigl[(d-2)(d-4)-(26-d)\bigr]}, \quad
d>d_{\text{cr}}. \label{45} 
\end{equation}
Because of the strong increase of the $\Gamma$ and $\zeta$ function in
the denominator, the left-hand side rapidly falls off for increasing
$d$. Typical values are $g^2 T_{\text{cr}}^{d-4}\simeq411.4, 12.0,
0.036$ for $d=8, 16, 26$. Therefore, the deconfined phase vanishes in
the formal limit $d\to\infty$.  

Incidentally, it is interesting to observe that the discontinuities of
the weight function $W(c)$ for $c\to0,1$ vanish for $d>5$; there,
$W(c)$ runs continuously to a finite extremal value for $c\to 0,1$. Between
four and five dimensions, the discontinuity persists and $W(c=0,1)$
increases for increasing $d$, finally approaching plus infinity at
$d\to5^-$. 

\section{Additional Infrared Scales in $d=4$}
\label{gluonmass}

The preceding section revealed that the $d=4$ model required
additional instructions on how to treat the singular infrared
modes. Although we rate the procedure established above as the most
general one of a ``universal'' character, we shall now suggest another
method, involving an additional scale. In the following investigation,
we exemplarily pick out one (physically motivated) possibility of
regularizing the infrared modes, and study its consequences. 

Let us assume that Yang-Mills theory dynamically generates a scale in
the infrared which can be reformulated in terms of an effective
mass\footnote{This mass should not be associated with a thermal gluon
  mass; the latter represents a collective excitation of the thermal
  plasma and is a typical feature of the high-temperature domain,
  being proportional to $T$. By contrast, the effective mass
  considered here shall particularly affect the low-temperature modes
  and be approximately constant in $T$.} $\meff$ for the transverse
fluctuating gluons\footnote{In this way, gauge invariance with respect
  to the background field is maintained.}. Although this scale may in
itself depend on some parameters, we shall consider it to be constant
within the limits of our investigation.

Adding the effective mass term to the inverse transverse gluon
propagator, e.g., in Eq.~\re{13}, it appears in a standard way in the
proper-time representation of the effective action; for example, the
integrand of the thermal one-loop contribution in Eq.~\re{30} is
multiplied by $\E^{-\meff^2 s}$ which damps away the infrared
singularities. Upon the substitution $u=\meff ^2 s$, we obtain
\begin{eqnarray}
\Leff^{1T}&=&-\frac{\meff^4}{4\pi^{2}} \tr_{\text{c}} 
\int\limits_0^\infty \frac{du}{u^{2}}\,\E^{-u} \frac{e_l}{\meff^2}
\left( \sinh \frac{e_l}{\meff^2} u + 
  \frac{1}{2\sinh \frac{e_l}{\meff^2} u} \right) \nonumber\\
&&\qquad\qquad\qquad\quad \times 
\sum_{n=1}^\infty \exp
  \left( -\frac{n^2}{4}\frac{\meff^2}{T^2} \frac{e_l}{\meff^2} \coth
  \frac{e_l}{\meff^2} u\right) \cos \frac{a_l}{T} n .
  \label{46}
\end{eqnarray}
Expanding for small ${e_l}/{\meff^2}$ in order to arrive at a
consistent derivative expansion for $A_0$, we find to order $e_l^0$
and $e_l^2$
\begin{eqnarray}
\Leff^{1T}\bigr|_0&=&-\frac{1}{\pi^2}\, \meff^2 \tr_{\text{c}}
\sum_{n=1}^\infty \frac{T^2}{n^2}\, K_2(\meff n/T) \, \cos
\frac{a_l}{T} n \equiv V(c,n^a,\meff), \label{47}\\
\Leff^{1T}\bigr|_{e_l^2}&=&-\frac{11}{24\pi^2} \, \tr_{\text{c}}\,
e_l^2 \sum_{n=1}^\infty K_0(\meff n/T) \cos \frac{a_l}{T} n
\nonumber\\
&&+ \frac{1}{24\pi^2} \, \tr_{\text{c}}\,
e_l^2 \sum_{n=1}^\infty \left( \frac{\meff}{T} n\right)  K_1(\meff
n/T) \cos \frac{a_l}{T} n, \label{48}
\end{eqnarray}
where we have employed a representation of the modified Bessel function
$K_\nu(x)$ \cite{GR}. Since we are interested in a possible formation
of a confinement phase, let us study Eqs.~\re{47} and \re{48} in the
low-temperature limit $T\ll \meff$. Then it is sufficient to use the
asymptotic form of the Bessel functions for large argument,
$K_\nu(x)\to \sqrt{\pi/2x} \E^{-x}$. 

Confining ourselves to the simplest case SU(2), we can deduce the
form of the potential from Eq.~\re{47}:
\begin{equation} 
V(c,\meff)\simeq-\sqrt{\frac{2}{\pi^3}}
T^4\left(\frac{\meff}{T}\right)^{3/2} \E^{-\meff/T}\left( \cos 2\pi c
  + \frac{1}{2} \right), \quad T\ll\meff. \label{49}
\end{equation}
Again, we encounter a potential with a (deconfining) maximum at
$c=1/2$, so that the effective mass does not induce significant
changes to the potential term.  

Including the contribution from the classical Lagrangian, the weight
function can be deduced from Eq.~\re{48} in the same limit:
\begin{eqnarray}
W(c,\meff)&=& T^2\left[ \frac{2\pi^2}{g^2} -\frac{1}{3}
  \sqrt{\frac{\pi}{2}} \left(\! 11\sqrt{\frac{T}{\meff}}
  -\sqrt{\frac{\meff}{T}} \right) \E^{-\meff/T} 
   \cos 2\pi c \right].
\label{50}
\end{eqnarray}
We first observe that, since the $\meff$ dependent term is
exponentially small for $T\ll \meff$, a small coupling $g^2$ will
always ensure that $W(c,\meff)$ is positive so that Polyakov loop
fluctuations are suppressed and the system is in the deconfined
phase. Therefore, the model predicts that confinement requires a
strong coupling. 

Indeed, if the coupling is (very) strong, we may neglect the first
term in Eq.~\re{50}, and find that $W(c,\meff)$ develops a minimum at
$c=1/2$, if\footnote{Taking the Bessel functions and the $n$ sum more
  accurately into account, the actual value of $T_{\text{cr}}$ changes
  slightly: $T_{\text{cr}}/\meff\simeq2/21$.}
\begin{equation}
T<T_{\text{cr}}, \quad
\frac{T_{\text{cr}}}{\meff}=\frac{1}{11}, \quad\text{for\,\,} g^2\gg
1. \label{51} 
\end{equation}
The situation can be rephrased as follows: if $T<T_{\text{cr}}$,
$c=1/2$ is the absolute minimum of $W(c,\meff)$. But $W(c=1/2,\meff)$
only becomes negative (thereby allowing for a confinement phase) if
the coupling is sufficiently large, so that the first term of
Eq.~\re{50} can be neglected.  

Therefore, our main conclusion of the present section is that a
different treatment of the infrared modes changes the behavior of the
model significantly! Although the present version of the model
exhibits the desired features, it requires more input and thus is less
meaningful: we need to specify the value of $\meff$ and the value of
$g^2$; the latter is involved with another scale $\mu$.

Let us end this section with the comment that the introduction of a
masslike infrared cutoff as employed in Eq.~\re{46} can also be used
as an alternative regularization scheme for the infrared modes. This
means that, giving up the meaning of $\meff$ as a physical scale, but
treating it as an arbitrary infrared cutoff scale for Eq.~\re{46}, we
may remove it after the calculation by taking the limit $\meff\to 0$
in Eqs.~\re{47} and \re{48}. We exactly recover Eq.~\re{32} (for
$d=4$), and, after analytical continuation, also Eq.~\re{36} with the
association $\meff\sim \mu$. The same procedure in $d>4$ dimensions
also leads to results identical with those in the preceding section.
It is in this sense that the treatment of the infrared modes as
performed in the preceding section can be rated ``universal''.

\section{Conclusions}
\label{comment}

In the present work, we have established and analyzed a dynamical
model for the order parameter of the deconfinement phase transition in
Yang-Mills theories -- the vacuum expectation value of the Polyakov
loop operator. We have calculated the effective action for this order
parameter to second order in a derivative expansion, and have treated
the gluonic fluctuations in one-loop approximation.

As a first conclusion, we observed that the ``constant-Polyakov-loop''
approximation, $A_0=$const., as considered in the literature, is in
principle incapable of describing two different phases owing to the
lack of an additional scale separating a high- and low-temperature
phase in $d=4$. This can also be inferred from the observation that
the vacuum expectation value of the trace of the energy momentum
tensor for a constant quasi-abelian $A_\mu$ background vanishes:
\begin{equation}
\langle T^\mu{}_\mu\rangle=\beta_{g^2}\, F_{\mu\nu}
F_{\mu\nu} =0,\quad \text{for\,\,}A_\mu^a=n^a A_\mu=\text{const.}
\label{52} 
\end{equation}
Therefore, a vacuum model of this type must necessarily preserve scale
invariance even at finite temperature so that the theory must remain
in a single phase. 

In the present model, scale-breaking is induced by fluctuations of the
Polyakov loop which, in a particular choice of gauge, are associated
with a nonvanishing electric field. The question of whether or not
these fluctuations are energetically favored can in principle be
answered by the dynamics of the model. In turns out that the
deconfinement phase is the generic phase in the absence of
fluctuations (this holds for all $d\geq 4$). Whether spatial Polyakov
loop fluctuations drive the model into a confining phase depends on
the form of the weight function $W(c)$ of the kinetic term. 

In four spacetime dimensions, thermal infrared singularities
complicate the investigation of the weight function and require
additional specifications of how to deal with these
singularities. Within a regularization-independent scheme that
introduces no other scale than already present, the $d=4$ model does
not reveal a confinement phase; instead, fluctuations of the
Polyakov loop even favor a deconfining vacuum state. 

By contrast, when regularizing the infrared by a physically effective
cutoff comparable to a gluon mass for the transverse modes, a phase
transition into a confining phase for low temperature becomes visible
in the strong-coupling regime. 

Whether one of these scenarios is realized in Yang-Mills theory
cannot, of course, be answered within a perturbative approach like
that employed in the present paper. Not only does the enormous
extrapolation of a one-loop calculation into the strong-coupling
sector present a major problem, but, with regard to the
infrared singularities, (even nonperturbatively) integrating out the
gluonic fluctuations at one fell swoop seems to be inappropriate.
Instead, the integration over the fluctuations should be performed
step by step in order to control a possible emergence of a dynamically
generated mass scale.

The one-loop model at least facilitates a concrete investigation of
possible scenarios, and at most displays some features in a
qualitatively correct manner. An appraisal of the different scenarios
requires further arguments. The first scenario of Sect.~\ref{4.1}
without an effective mass can be preferred only from a theoretical
viewpoint owing to its simplicity and universality. Though the second
scenario of Sect.~\ref{gluonmass} needs more input, the appearance of
an additional infrared mass scale is common to almost all conjectured
confining low-energy effective theories of Yang-Mills theory;
therefore, a phenomenological viewpoint supports this scenario from
the beginning, and so does the final result. Nevertheless, a reliable
investigation of the infrared requires nonperturbative methods.
 
Let us finally comment on the differences of our results to
Ref.~\cite{enge98} which inspired the model considered in the present
work; although the representations of the effective action in the form
of Eq.~\re{6} are congruent, the meaning of the results is quite
different in comparison: in \cite{enge98}, the fluctuations of the
$A_0$ field have not been taken into account, implying that the
resulting ``effective action'' remains a complete quantum
theory of the $A_0$ field. The $A_0$ ground state is then {\em
  approximated} by the effective potential which is obtained by
transforming the kinetic term to standard canonical form. By contrast,
we integrated over {\em all} quantum fluctuations of the $A_\mu$ field
in the present work; therefore, the resulting effective action is the
generating functional of the 1PI diagrams and governs the dynamics of
the background fields in the sense of classical field theory. To
conclude, it is not astonishing that the explicit results of
\cite{enge98} in particular for the weight function $W(c)$ do not
agree with ours because they have a different origin and a different
meaning.\footnote{From \cite{enge98}, it is in principle possible to
  arrive at our results (and get rid of the renormalization problems)
  by integrating over the $A_0$ fluctuations; however, for a
  consistent treatment, the second weight function of the ${\cal
    O}((\partial c)^4)$ term has to be known before integrating over
  the $A_0$ fluctuations.}

In spacetimes with more than four dimensions, the situation simplifies
considerably: on the one hand, the model is infrared finite, thereby
producing unambiguous results; on the other hand, there already exists
another dimensionful scale given by the coupling constant. We
discovered a phase transition from the (generic) deconfining to a
confining phase for {\em increasing} temperature for
$d>d_{\text{cr}}\simeq7.42$. This is consistent with the fact that the
dimensionless coupling constant grows for increasing energies,
reaching an UV-stable fixed point. Beyond perturbation theory, the
latter statement has also been confirmed in the nonperturbative
framework of exact renormalization group flow equations \cite{reut94}.

\section*{Acknowledgments}
The author wishes to thank W. Dittrich for helpful conversations and for
carefully reading the manuscript. Furthermore, the author profited
from insights provided by M. Engelhardt, whose useful comments on the
manuscript are also gratefully acknowledged.

\end{document}